\documentclass{stanfest}
\usepackage{amsmath}
\usepackage{amssymb}

\newcommand{\unit}{1\!\!1}

\def\vel{\, \mathbf v}
\def\Bfield{\, \mathbf B}
\def\Efield{\, \mathbf E}
\def\Fpart{\, \mathbf F_{\rm part}}
\def\upart{\, \mathbf u_{\rm part}}
\def\qpart{\, q_{\rm part}}
\def\Jpart{\, \mathbf J_{\rm part}}

\author{Allard Jan van Marle}[ELI, LUPM]

\affil[ELI]{ELI Beamlines, Dolní Břežany, Czech Republic}
\affil[LUPM]{LUPM, Montpellier, France}

\title{Combining PIC and MHD to model particle acceleration in astrophysical shocks}
\headtitle{modelling particle acceleration with PIC-MHD}
\begin{document}

\maketitle

\begin{abstract}
When supersonic plasma flows collide, many physical processes contribute to the morphology of the resulting shock. One of these processes is the acceleration of non-thermal ions, which will, eventually, reach relativistic speeds and become cosmic rays. This process is difficult to simulate in a computer model because it requires both macro-physics (the overall shape of the shock) and micro-physics (the interaction between individual particles and the magnetic field).

The combined PIC-MHD method is one of several options to get around this problem. It is based on the assumption that a plasma can be described as a combination of a thermal gas, which can be accurately described as a fluid using grid-based magnetohydrodynamics (MHD) and a small non-thermal component which has to be described as individual particles using particle-in-cell (PIC).

By combining aspects of both methods, we reduce the computational costs while maintaining the ability to trace the acceleration of individual particles. We apply this method to a variety of astrophysical shock configurations to investigate if, and how, they can contribute to the cosmic ray spectrum.

\end{abstract}

\section{Introduction}
\label{sec:intro}
Astrophysical shocks are widely considered the primary source of cosmic rays (CRs), which are charged particles that have been accelerated to relativistic speeds. The acceleration mechanism involves repeated shock crossing, with the particle gaining momentum as it is reflected across the shock through interaction with the local magnetic field \citep[e.g.][]{Bell78,BlandfordOstriker78,Drury83}. This process not only changes the particles' momentum but also the shock structure. Energy gained by the particles is drained from the thermal plasma of the shock, and the presence of a charged particle current can trigger instabilities both up- and downstream of the shock front, which can result in distortions of the shock.

However, although it is a widely recognized part of shock physics, particle acceleration is rarely considered in numerical models of astrophysical shocks. One of the main reasons for this is the numerical challenge of modelling a process involving several orders of magnitude in both space and time \citep[e.g.][]{Marcowith16}. 

The most commonly used method to simulate shocks is magnetohydrodynamics (MHD), which treats the gas as a compressible fluid and determines its characteristics by solving a set of conservation equations on a grid. The underlying assumption is that within each grid-cell, the gas can be represented by a handful of physical parameters: mass density, momentum density, energy density and magnetic field strength. The main advantages of this method are numerical. Numerical MHD is computationally efficient and can be scaled well to model large structures by either increasing the size of each grid-cell or increasing the number of cells. Unfortunately, numerical MHD also has limitations, which lie, primarily, in the physics in can include. Since the gas is described based on thermal averages, the gas has to be in local thermodynamic equilibrium. Otherwise, the averages would not be representative. Similarly, numerical MHD cannot be used to model the micro-physics of particles because their individual behaviour is eliminated by the fluid approximation.

An alternative method for simulating shocks is particle-in-cell (PIC). This method also uses a grid to define the spatial domain, but treats the plasma as a collection of particles, representing the ions that form the plasma. Rather than solving conservation equations, this method treats each particle kinetically and the Maxwell equations to derive the electromagnetic field properties which, in turn, exert a force on the particles. Because PIC treats the plasma as a collection of particles, it has no trouble modelling a plasma that is not in equilibrium and particle physics is, inherently, part of the method. However, these characteristics come at a computational cost. PIC is inherently computationally inefficient because a large number of particles are required in each grid cell to provide a statistically significant sample. Furthermore, PIC is prone to numerical noise, which requires additional computation to eliminate. As a result, PIC will quickly run into trouble when modelling large-scale structures.

When comparing the two methods (See Table~\ref{tab:characteristics}), it becomes clear that the ideal solution would be to find a combination of the two, which plays to the strengths of each method, while allowing them to compensate for each other's weaknesses. 
There are several ways to do this, including: Spatial separation (with part of the spatial domain treated with MHD and others with PIC); particle species separation (commonly known as di-hybrid), which treats ions as particles and electrons as a fluid); and PIC-MHD.

\subsection{PIC-MHD}
\label{sec-picmhd}
For PIC-MHD \citep{Baietal:2015,vanMarleetal:2018, Amano18, Mignone18}, the separation is not based on either particle species or spatial location, but on particle characteristics. PIC-MHD assumes that the plasma can be separated into two components: thermal and non-thermal. 
This being the case, the plasma can be described as a primarily thermal plasma, (simulated through MHD), which has a non-thermal correction term (supplied by PIC). 
Both thermal and non-thermal components exist in the same time and space and interact with each other. 
To accomplish this, the PIC-MHD method utilizes a single grid that functions as the computational grid for both the MHD code and the PIC code. Thermal gas quantities are defined at the cell centres, which double as corners for the PIC code. In terms of physics, the conservation equations have to be adapted to include the presence of a non-thermal plasma component.

The PIC-MHD method has several advantages. Contrary to pure MHD models, it can simulate the behaviour of particles. However, unlike PIC, it does not require a large particle population because the thermal plasma is modelled using thermal averages. This makes the PIC-MHD computationally efficient without sacrificing the ability to simulate microphysics. 

Inevitably, the PIC-MHD method also has limitations. In its current form, it requires a primarily thermal plasma. If the non-thermal component becomes too large (more than a few percent of the total number density), relative to the thermal plasma, the underlying assumptions of the PIC-MHD model break down. However, this is not a major restriction since many astrophysical plasmas fit within these limitations
The main downside is that PIC-MHD lacks a physical description of the transition of a particle from thermal to non-thermal. When the thermal plasma passes through a shock, a fraction of the particles become non-thermal. However, the PIC-MHD method does not include the physics required to model this process because the shock is treated as an MHD discontinuity.  Instead, it relies on an ad-hoc model that takes a given fraction of the thermal plasma, removes it from the thermal gas and re-injects it as non-thermal particles.

\begin{table}[!t]

\caption{Characteristics of MHD and PIC compared}
\label{tab:characteristics}
\begin{center}
{\small
\begin{tabular}{lcc}
\hline
{\bf } & {\bf MHD} & {\bf PIC} \\
\hline
{\bf Efficiency} & good & poor\\
{\bf Scaling } & good & poor\\
\hline
{\bf Non-thermal gas} &  no & yes\\
{\bf Micro physics} &  no & yes\\
\hline

\hline
\end{tabular}
}
\end{center}

\end{table}

\begin{figure}
\centering
\mbox{
\includegraphics[width=0.6\textwidth]{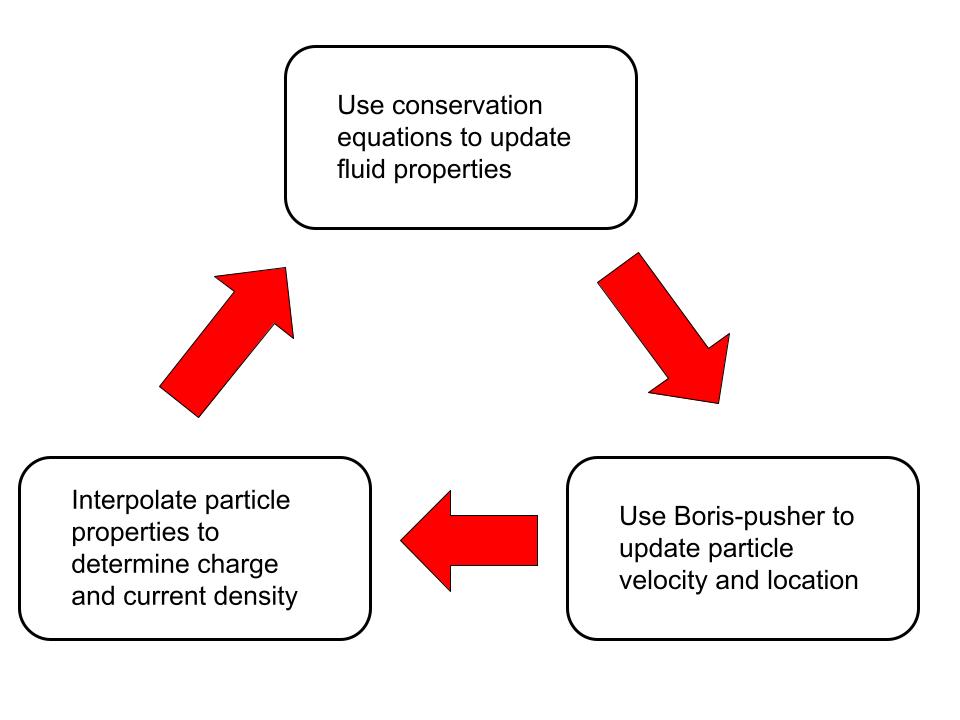}}
\caption{The computational loop of the PIC-MHD model}
 \label{fig:PICMHD_diagram}
\end{figure}

\section{Numerical implementation}
\label{sec-numerics}
In \citet{vanMarleetal:2018}, henceforth paper~1, van Marle, Casse \& Marcowith introduced a code, implementing the PIC-MHD method. This code uses the {\tt MPI-AMRVAC} code \citep[e.g.][]{vanderHolstetal:2008} as a basis. This state-of-the-art, fully conservative MHD code solves the conservation equations on an adaptive mesh grid and incorporates a large variety of numerical solvers. van Marle, Casse \& Marcowith implemented a new module that solves both the movement of particles and the interaction between the thermal and non-thermal components of the plasma.

The presence of the charged, non-thermal particles creates a current. This in turn generates an electric field component. As a result, Ohm's law, which connects the electric field $\Efield$ to the magnetic field, $\Bfield$ has to be rewritten. As described in \citet{Baietal:2015}, the new Ohm's law takes the form
\begin{equation}
c\Efield = -\left((1-R)\vel +R\upart\right)\times\Bfield,
\label{eq:ohms}
\end{equation}
with $c$, the speed of light, $\vel$ and $\upart$ the velocity of the thermal and non-thermal plasma components respectively, and $R$ is the supra-thermal particle charge density relative to the total charge density. Both $\upart$ and $R$ can be obtained by mapping the contributions of individual particles onto the cell centres of the grid.
N.B This form of Ohm’s law neglects the contributions of the Hall current and the electron thermal plasma. These terms can be safely discarded provided the magnetic field pressure is non-negligible compared to thermal pressure and typical length scales are larger than the ion skin depth. We also assume that in each grid cell the combined thermal and non-thermal plasma is always neutral.

\subsection{Conservation equations}
Having re-written the Ohm's law (Eq.\ref{eq:ohms}), we can obtain a new version of
the MHD conservation equations for mass, momentum and energy, including the additional terms that arise from the interaction with the non-thermal particles are. The mass conservation equation remains unchanged (We do not, at this time, include particles transitioning between the thermal and non-thermal components):
\begin{equation}
\frac{\partial \rho}{\partial t} ~+~ \nabla \cdot (\rho \vel)~=~0,
\label{eq:mass}
\end{equation}
with $\rho$ and $\vel$ the mass density of the thermal plasma. 
The momentum equation has to be changed to include an extra force term, representing the force that allows the electromagnetic field to influence the motion of the particles:
\begin{equation}
\frac{\partial \rho\vel}{\partial t} + \nabla\cdot\left(\rho\vel\otimes\vel-\frac{\Bfield\otimes\Bfield}{4\pi}+P_{\mathrm tot}\unit\right)~=~-\Fpart,
\label{eq:momentum}
\end{equation}
with $P_\mathrm{tot}=P+B^2/8\pi$ the total pressure and $\Fpart$ the force density applied by the thermal plasma onto the non-thermal particles. 
This extra force term requires an extra work term in the energy equations, which becomes:
\begin{equation}
\frac{\partial e}{\partial t}+\nabla\cdot\biggl( (e + P_{\rm tot})\vel+(\Efield-{\mathbf E}_0)\times\frac{\Bfield}{4\pi}\biggr) 
~=~ -\upart \cdot \Fpart
\label{eq:energy}
\end{equation}
with $e$ the total energy density of the thermal plasma and $E_0$ the electric field that the thermal plasma would have if no non-thermal particles were present.

\subsection{Particle motion}
The motion of the particles is determined by the force exerted by the electromagnetic field on the particles: 
\begin{equation}
\Fpart~=~(1-R) \biggl( \qpart  {\mathbf E}_0 + \frac{\Jpart}{c} \times \Bfield \biggl), 
\end{equation}
with $\qpart$ and $\Jpart$ the non-thermal particle charge and current densities.

\subsection{Numerical process}
In practice, the PIC-MHD code starts each time step by mapping the particle charge- and current-densities, obtained from the distribution of the particles, onto the centres of the grid cells. Once this is done, it solves the conservation equations, advancing the characteristics of the thermal plasma by a single time step. Using the initial values of the electric and magnetic field strengths, it then calculates the change in location and momentum for each non-thermal particle using a Boris method \citep[][and references therein]{BirdsallLangdon:1991}. 
Once that is done, both the thermal and non-thermal components have now moved one step forward in time. and the process can begin again (See Fig.~\ref{fig:PICMHD_diagram}).

\section{Numerical results}
\subsection{2-D parallel shocks}
\label{sec-2D}
Paper~1 showed the initial results for a 2-D model of a parallel shock, including particle acceleration. This simulation was set up to resemble earlier simulations by \cite{Caprioli14a,Caprioli15} and \cite{Baietal:2015} and served primarily as a test case for the code. 

The simulation used a box of 240$\times$30  times a reference gyro radius $r_{\rm g}={m \,v_{\rm inj}}/q B_0$, which corresponds to the gyro radius of particles with charge $q$ and mass $m$ inserted with an initial velocity $v_{\rm inj}$ in the undisturbed upstream magnetic field of strength $B_0$.  
At the lowest resolution level, each grid-cell was $1\times\,1\,r_{\rm g}$, with the adaptive process using four mesh levels of refinement depending on the MHD physical conditions throughout the domain. As a result, our grid could reach a maximum effective resolution of $3840\times480$ cells. \\

Unlike more traditional PIC and PIC-hybrid models, which form a shock by letting a plasma flow collide with a fixed wall, the PIC-MHD method can run in the frame of reference of the shock. Therefore, the shock is effectively stationary, which reduces the required box size. 
For this particular model, we chose an initial shock velocity of $0.003c$ and a Mach number of 30, assuming that the pre-shock plasma is in equipartition (magnetic and thermal energy being equal). We also assumed that only protons become non-thermal, which removed the problem of resolving both electron and proton length and time scales. 
As discussed in Sect.~\ref{sec-picmhd}, the PIC-MHD method does not provide a transition between the thermal and non-thermal gas. Therefore, the injection of particles at the shock had to be parametrized. Here we chose to emulate the setup by \citet{Baietal:2015}, which assumed an injection fraction of $2.10^{-3}$ of the mass crossing the shock. This mass was removed from the thermal plasma and re-injected in the form of particles with the characteristics of protons, moving at 3 times the shock velocity  with an isotropic velocity distribution in the down-stream medium.  

\begin{figure*}
\centering
\mbox{
\includegraphics[width=0.45\textwidth]{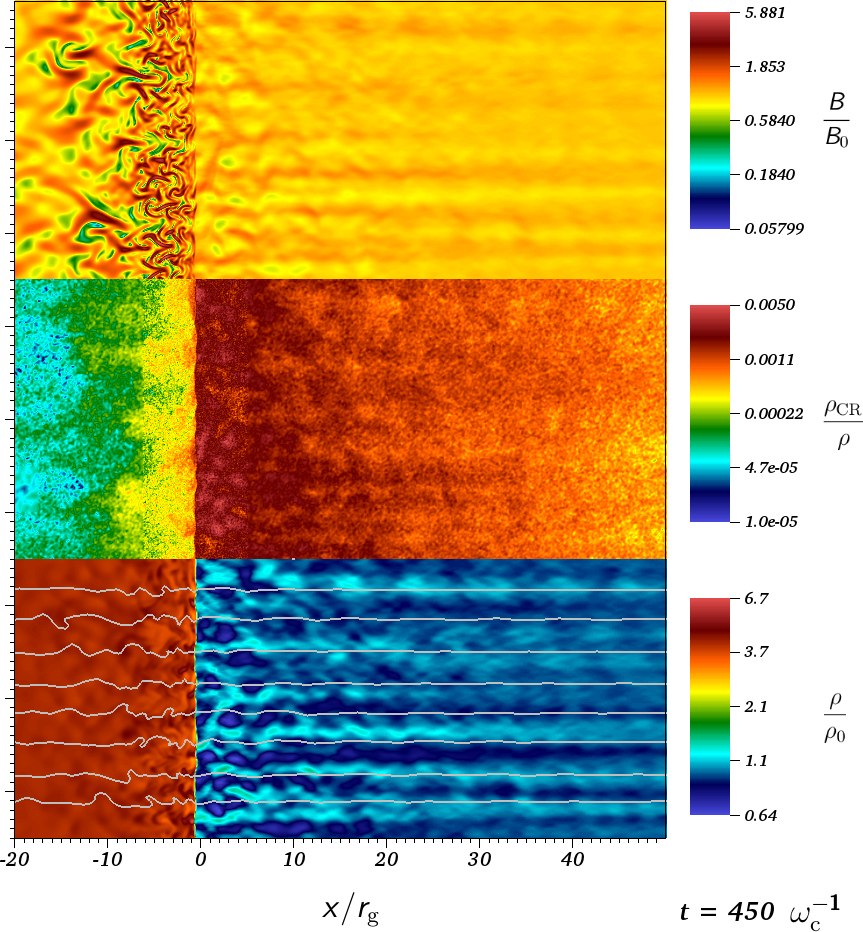}
\includegraphics[width=0.45\textwidth]{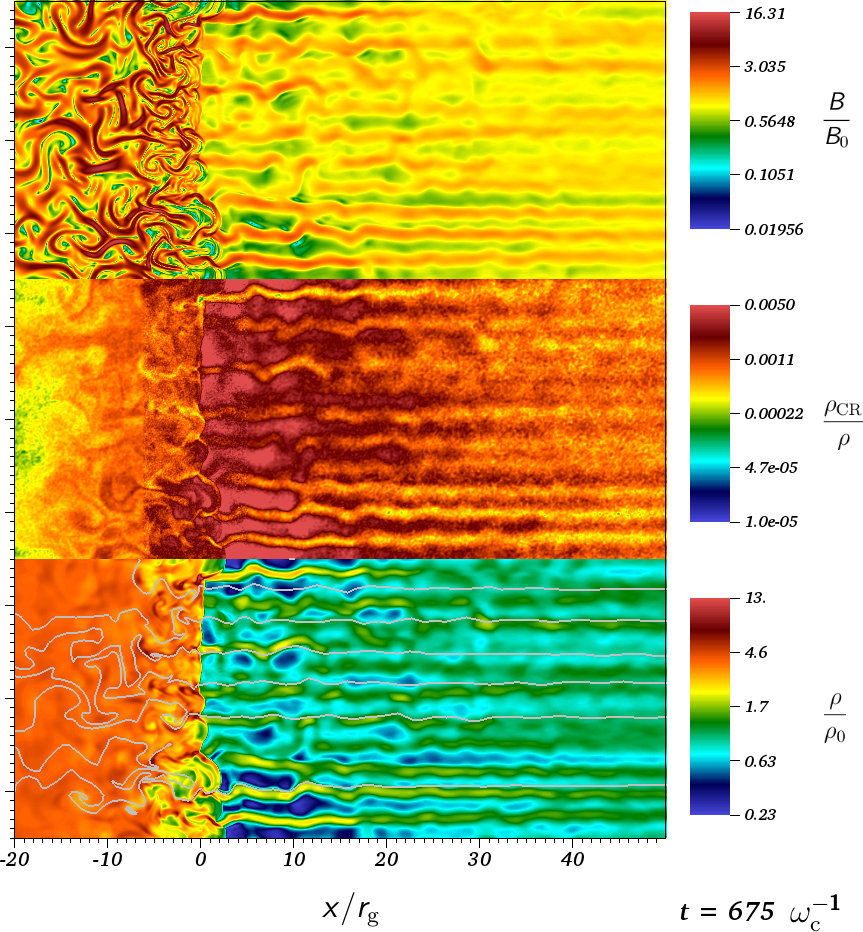}}
\caption{From top to bottom, each panel shows the magnetic field amplitude, cosmic-ray density and plasma density of the 2-D parallel shock simulation, combined with the magnetic field stream lines. Over time, the filamentary structure that indicates the non-resonant streaming instability becomes more pronounced and causes distortions in the shock front.   [Reproduced with permission from van \citet{vanMarleetal:2018}. Copyright 2018 OUP Publishing.]}
 \label{fig:2Devolution}
\end{figure*}

The result of this simulation is shown in Fig.~\ref{fig:2Devolution}, which shows two time-frames. In the upstream medium, the magnetic field and the thermal gas both show a filamentary structure, parallel to the magnetic field. This is characteristic of the non-resonant streaming instability  \citet{Bell04}. Downstream of the shock, the medium is more randomly turbulent. This behaviour becomes more extreme over time, to the point where the shock front becomes distorted owing to the asymmetry in the ram pressure it experiences.

These results closely resemble the analytical results from \citet{Bell04} as well as the numerical results obtained by \citet{Reville08, Reville13}, which used the Vlasov-Fokker-Planck (VFP) method, \citet{Caprioli14a,Caprioli15} using the di-hybrid method, as well as the PIC-MHD results obtained by \citet{Baietal:2015}, proving the validity of our code.

\subsection{3-D parallel shocks}
\begin{figure*}
\centering
\mbox{
\includegraphics[width=0.45\textwidth]{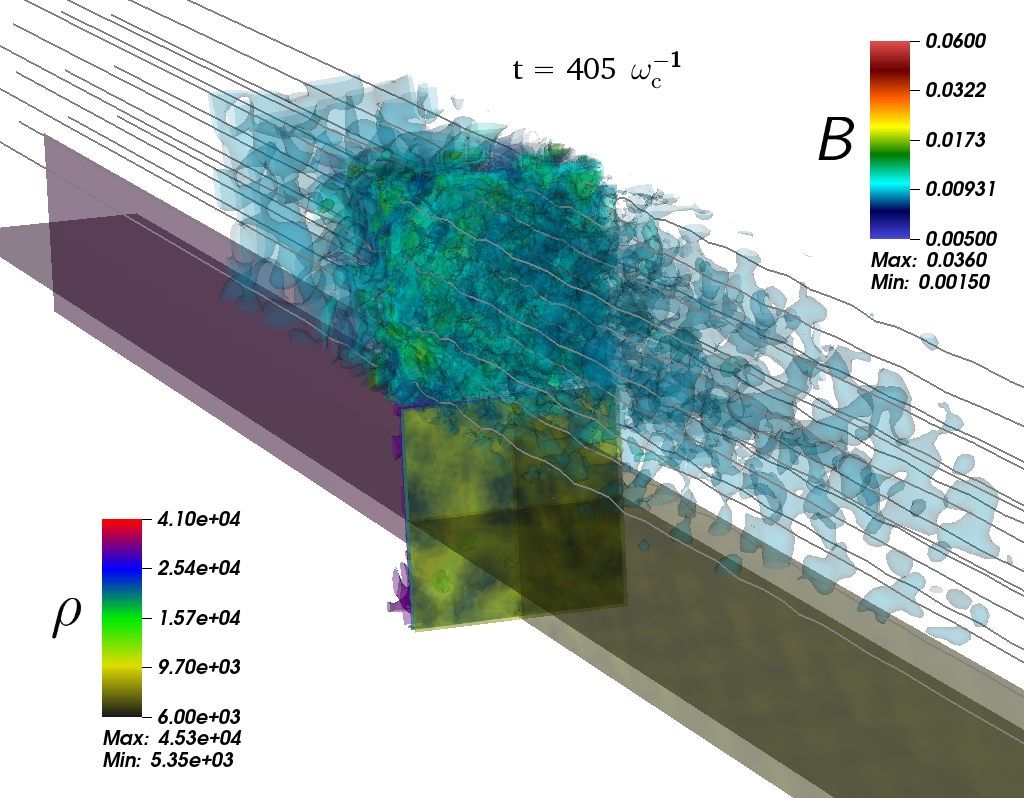}
\includegraphics[width=0.45\textwidth]{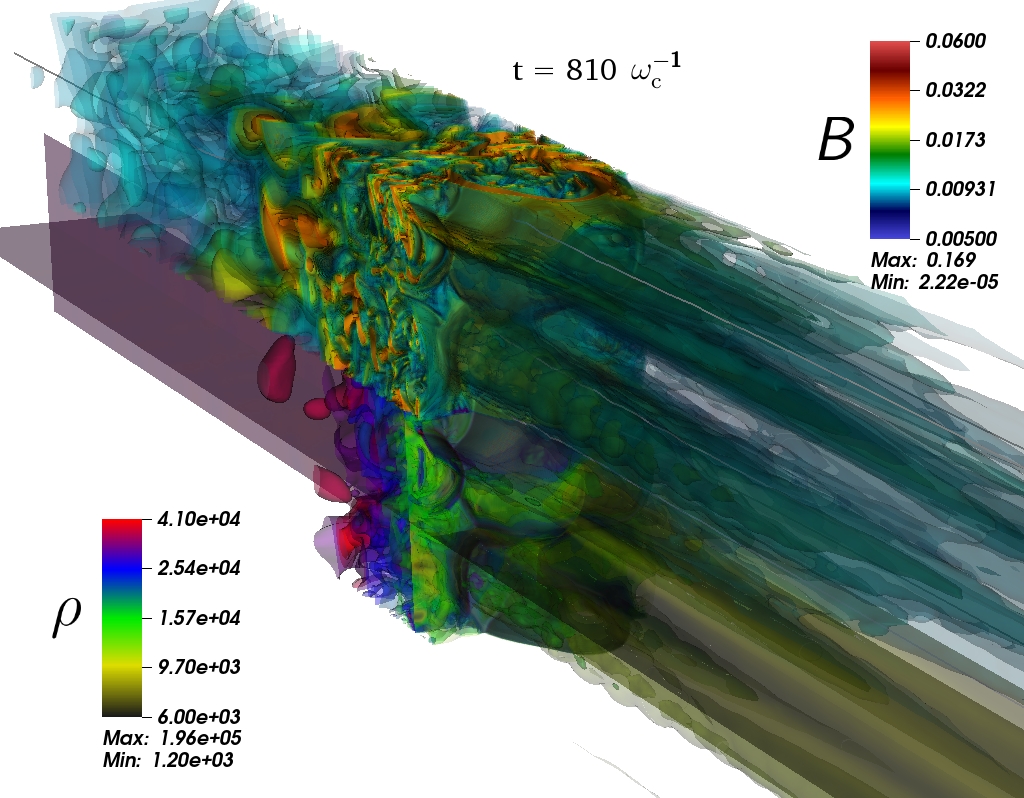}}
\caption{Magnetic field strength and thermal plasma density for a 3-D simulation Initially, (t=405 $\omega_{\rm c}^{-1}$, left panel) the distortion in the magnetic field appears random and the thermal gas density remains undisturbed. However, over time (t=810 $\omega_{\rm c}^{-1}$, right panel) the disturbance in the upstream magnetic field and thermal gas density demonstrate the characteristic filaments of the non-resonant streaming instability. The downstream magnetic field becomes highly turbulent and the shock front becomes highly distorted. [Reproduced with permission from van \citet{vanmarleetal:2019}. Copyright 2019 OUP Publishing.]}
 \label{fig:3Devolution}
\end{figure*}

Paper~1 showed the initial results for a 2-D model of a shock, including particle acceleration. With these results in mind, we used the computational efficiency of the PIC-MHD method to extend our models to 3-D, the results of which were published in \citet{vanmarleetal:2019}. 
We used the same physical setup as in Paper~1, albeit the box size had to be reduced to save computational time. Therefore, we limited the spatial domain to $192\times12\times12\,r_{\rm g}$. Grid size started with a basic grid of $640\times40\times40$ cells, with two additional levels of refinement allowed to give us an effective grid of $2560\times160\times160$ cells. 
The results, as shown in Fig.~\ref{fig:3Devolution} show the upstream filamentary structure in 3-D, as well as the distortion of the shock front. This proves that these are physical features and not artefacts of the 2-D slab symmetry of previous simulations

\subsection{Quasi-perpendicular shocks}
So far, the models shown only pertained to parallel shocks, where the magnetic field is aligned with the flow. Initially, we investigated quasi-perpendicular shocks by repeating the 2-D model (Sect.~\ref{sec-2D}) but with the magnetic field at an angle with the flow. The results, presented in Paper~1, indicated that such shocks could still accelerate particles, in contradiction with earlier results by \citet{Caprioli14a,Caprioli15}. This could be partially explained by a longer simulation time and a larger spatial domain, which allowed us to capture long-wavelength instabilities.  However, the PIC-MHD models in Paper~1 assumed a similar injection fraction to that of a parallel shock, which is not necessarily realistic \citep{Haggerty:2019}. 
In \citet{vanMarleetal:2022}, we introduced a more realistic model that combined the PIC-MHD 2-D simulations with a reduced particle injection rate calculated through a PIC model of the shock structure. 
The results showed that quasi-perpendicular shocks can only accelerate particles if the Alfv{\'e}nic Mach number is sufficiently high. Otherwise, the magnetic field is too strong for the particle current to destabilize.

\section{Outlook}
The PIC-MHD method has demonstrated that it is a viable alternative to the PIC and di-hybrid methods, allowing for plasma simulations that include particle physics to be run on relatively small computer architectures. (The 2-D models shown in this paper were run on desktop workstations, the 3-D model was performed on a Tier~2 cluster.) 
The use of traditional, grid-based MHD also enables the user to experiment with a wide range of input parameters. E.g. \citet{vanmarle:2020} explored the behaviour of a high-$\beta$, low-Mach shock, such as galaxy cluster collision shocks, demonstrating the ability of these shocks to accelerate particles. 
Meanwhile, \citet{Plotnikovetal:2024} included PIC-MHD models that simulated the interaction between an electron-positron flow and a thermal plasma, such as may occur in pulsar-wind nebulae.

We are currently in the process of extending our models to include new physics, such as special relativistic MHD (SRMHD), which allows us to model relativistic flows and shocks. Initial results of this method were shown in \citet{2023arXiv230812721B,2024arXiv:2407.05847}. 
Future developments will also include a new version of Ohm's law, allowing us to model plasmas with a larger non-relativistic component than is currently the case. 

Finally, the PIC-MHD method can be extended to simulate the behaviour of electrically charged dust grains, rather than ions and electrons. Though the scaling is very different, the principles remain the same. This would enable us to accurately model the dust distribution in objects such as molecular clouds and proto-planetary disks, and match the results to radio and infrared observations.

%
\acknowledgements{
With special thanks to Alexandre~Marcowith, Fabien~Casse, Artem~Bodahn, Illya Plotnikov, Anabella Araudo, Martin Pohl, Paul Morris, Claire Gu{\'e}pin, and Pierrick Martin
 for years of collaboration. The simulation results shown in this paper were originally published in {\tt MNRAS}.
This work made extensive use of NASA's Astrophysics Data System (ADS).}

%
\bibliographystyle{stanfest_bibstyle}
\bibliography{vanmarle_references}

\end{document}